\documentclass[twocolumn,aps,prb,showpacs,amsfonts,amssymb,amsmath]{revtex4}
\usepackage{mathrsfs}
\usepackage{bm}
\usepackage{amsmath}
\usepackage{amssymb}
\usepackage{graphicx,times,graphics,color,epsfig}

\begin{document}
\draft
\title{The statistics of Wigner delay time in Anderson disordered systems}

\author{Fuming Xu}
\author{Jian Wang}
\email{ jianwang@hkusub.hku.hk}

\address {Department of Physics and the Center of Theoretical and
Computational Physics, The University of Hong Kong, Hong Kong,
China}

\begin{abstract}
We numerically investigate the statistical properties of Wigner
delay time in Anderson disordered 1D, 2D and quantum dot (QD)
systems. The distribution of proper delay time for each conducting
channel is found to be universal in 2D and QD systems for all
Dyson's symmetry classes and shows a piece-wise power law behavior
in the strong localized regime. Two power law behaviors were
identified with asymptotical scaling ${\tau^{-1.5}}$ and
${\tau^{-2}}$, respectively that are independent of the number of
conducting channels and Dyson's symmetry class. Two power-law
regimes are separated by the relevant time scale $\tau_0 \sim
h/\Delta$ where $\Delta$ is the average level spacing. It is found
that the existence of necklace states is responsible for the second
power-law behavior ${\tau^{-2}}$, which has an extremely small
distribution probability.
\end{abstract}

\maketitle

\section{introduction}

After the pioneering works,\cite{eisenbud,wigner,smith} the problem
of quantum mechanical scattering has attracted intensive research
interest in many fields. The dynamical aspect of the scattering
process can be characterized by the energy derivatives of the
scattering matrix, known as the Wigner delay time $\tau$.
Semi-classically, Wigner delay time measures the time spent by the
center of a wave packet in the scattering region, which is simply
related to its group velocity. Since $\tau$ is critically dependent
on the transport process, it is not
self-averaging\cite{avering1,avering2}. Hence a complete
distribution is required to fully comprehend this quantity in
classical chaotic or quantum disordered systems.

The distribution of delay time in 1D system has been thoroughly
studied both theoretically
\cite{comet,joshi,texier,stein,ossipov1,kumar,amrein,bodyfelt} and
experimentally.\cite{genack,chananov} Texier and Comet\cite{texier}
showed by various methods the universality of the distribution of
$\tau$ in a 1D semi-infinite system, which has a ${\tau^{-2}}$
power-law tail at large $\tau$ in the localized regime. This power
law tail has also been confirmed by others.\cite{joshi,stein,kumar}
It has been established that ${\tau^{-2}}$ behavior is valid for
different types of disorder potential with $\delta$, gaussian, box,
and exponential decaying distributions.\cite{comet,joshi}

For 2D systems with multichannels, most of the theoretical works are
based on the random matrix theory
(RMT).\cite{fyodorov1,seba,brouwer,fyodorov2,sommers} A simple
expression of $P_N(\tau)$ for different Dyson's symmetry classes has
been derived\cite{seba,fyodorov1} at the ideal coupling condition,
which has a $\chi^2$ distribution with $\beta(N+1)$ degrees of
freedom where $\beta$ is the symmetry index and N the number of
conducting channels. On the other hand, a $P(\tau) \sim \tau^{-1.5}$
asymptotic behavior in the diffusive regime was numerically obtained
from the kicked rotor model (KR
model).\cite{ossipov2,kottos1,kottos2} Starting from RMT, it was
shown that an open chaotic system weakly coupled with many open
channels also exhibits the $\tau^{-1.5}$ power-law behavior in
symmetry classes $\beta$=1, 2, and 4.\cite{fyodorov2,sommers} So far
all the studies are focused on diffusive regime, less attention has
been paid on the localized regime. It is the purpose of this paper
to fill this gap. In this paper, we wish to explore the possibility
of universal behavior of the distribution of Wigner delay time
$\tau$ in the localized regime.

\begin{figure}[tphp]
\centering
\includegraphics[width=9cm]{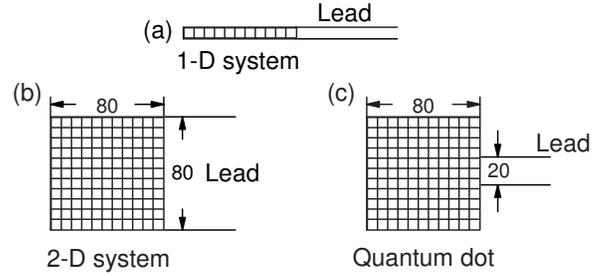}
\caption{sketch of the three geometries of interest. (a) 1D
semi-infinite tight-binding chain; (b) 2D system; (c) QD
system.}\label{fig1}
\end{figure}

In this paper, we carry out an extensive numerical investigation of
statistical properties of Wigner delay time $\tau$ in disordered
systems with Anderson-type of impurities for 1D, 2D and quantum dot
(QD) systems. For 1D systems, our results confirm that the
distribution of Wigner delay time follows a power-law behavior
${\tau^{-2}}$ in the localized regime. For 2D and QD systems, our
results show that the distribution of proper delay time of each
conducting channel obeys a universal piece-wise power-law in strong
localized regime that is independent of number of channels and
Dyson's symmetry index $\beta$. For a scattering system, the
characteristic time scale $\tau_0$ is related to the group velocity
$v_g$ of the electron, i.e., $\tau_0 = L/v_g$ where $L$ is the
characteristic length of the system. Our results show that when
$\tau < \tau_0$, the distribution follows a power-law
${\tau^{-1.5}}$ while for $\tau
> \tau_0$ a new power-law of ${\tau^{-2}}$ is obtained for the
proper delay time. Our result indicates that the power-law of
${\tau^{-2}}$ can only be observed in the localized regime. The
physical origin of the new power-law behavior is the existence of
the so called Azbel resonant state\cite{azbel1,azbel2} or necklace
state\cite{necklace1,necklace2}. This necklace state has a very long
lifetime and is the multi-resonant state inside the scattering
region in the localized regime. When the incoming electron has N
conducting channels, the distribution for total delay time is given
by $P(\sum_i \tau_i)$. Although $P(\tau_i)$ obeys the piece-wise
power-law behavior for each proper delay time $\tau_i$, the
distribution for total delay time behaves differently. Our results
show that as $N$ increases the power-law region for ${\tau^{-1.5}}$
becomes narrow and eventually diminishes in the large N limit while
the power-law region for ${\tau^{-2}}$ remains but shifts towards
small $\tau$. Therefore, in the large N limit, only one power-law
scaling of ${\tau^{-2}}$ survives in the localized regime. This
conclusion is valid for 2D and QD systems and for three different
ensembles with $\beta=1,2,4$.

This paper is organized as follows. In section II, a theoretical
formalism and a numerical implementation scheme for calculating
Wigner delay time are given. In section III, extensive numerical
results and analysis for distribution of Wigner delay time are
presented for 1D, 2D and QD systems with different symmetries
($\beta=1,2,4$). Finally a brief summary is given in section IV.

\section{theoretical formalism}

Following Wigner\cite{wigner} and Smith,\cite{smith} the
Wigner-Smith delay time matrix is defined in terms of the scattering
matrix S as
\begin{eqnarray}
Q(E)=-i\hbar S^{\dag}(E) \frac{\partial S(E)}{\partial E} \nonumber
\end{eqnarray}
The delay time $\tau$ is simply the summation of the diagonal
elements of matrix Q(E)
\begin{eqnarray}
\tau=Tr[Q]=-i\hbar Tr[S^{\dag} \frac{\partial S}{\partial E}]
\label{e1}
\end{eqnarray}
From now on we will omit the energy dependence of the relevant
quantities for simplicity. Supposing there are N conducting
channels, the eigenvalues of delay time matrix Q are called the
proper delay times, $\tau_1, \tau_2, ..., \tau_N$, which can be
viewed as the contribution to the total delay time $\tau$ from the
corresponding conducting channel.

Three setups under investigation are schematically illustrated in
Fig.\ref{fig1}, which are respectively (a) a 1D semi-infinite chain
with length $L=1000 a$ lattice points; (b) a 2D system with 80 by 80
lattice sites connected to a single lead with width $W_0=80a$; (c) a
quantum dot system with the same number of lattice points as that of
the 2D system but with a lead of narrower width $W_0=20a$. Here $a$
is the lattice spacing between two adjacent sites, being the length
scale in the calculation. Most of numerical calculations were done
using these parameters. In the present work we use the conventional
nearest-neighbor tight-binding approximation and Green's function
formalism to numerically study these systems. All these geometries
are connected by only one semi-infinite lead to the electron
reservoir, which ensures the unitary of the scattering matrix since
all electrons incident will be reflected back into the reservoir in
these one-lead systems.

Assuming our 2D system is in the x-y plane. In the presence of
perpendicular magnetic field and Rashba spin-orbit interaction, the
generalized Hamiltonian is given by
\begin{eqnarray}
H=\frac{1}{2 m^*} [p + \frac{e\textbf{A}}{c}]^2 + V +
{\mu}B{\cdot}{\sigma} + \frac{t_{SO}}{\hbar} [\sigma \times (p+
\frac{e\textbf{A}}{c})]_z \label{e2}
\end{eqnarray}
where $p$ is the momentum and $m^*$ the effective mass of electron.
Here $V$ is the confining potential that is set to zero inside the
device and infinity at the boundary of the device except at the
interface of the lead. The vector potential \textbf{A} due to the
magnetic field is expressed as \textbf{A}=(-By, 0, 0) under Landau
gauge with B the magnetic field. $\sigma$ is the Pauli matrix and
$t_{SO}$ is the strength of spin-orbit coupling. $\mu=g \mu_B/2$ is
the magnetic moment, with $g=4$ the Lande g-factor and $\mu_B$ the
Bohr Magneton. The tight-binding Hamiltonian with nearest-neighbor
hopping has an expression\cite{qiao}
\begin{equation}
\begin{array}{cll}
H=& \sum_{nm\sigma}(\varepsilon_{nm} +
\epsilon_{nm})c^\dag_{nm\sigma}
c_{nm\sigma} \\
& - t\sum_{nm\sigma}[ c^\dag_{n+1,m\sigma} c_{nm\sigma}
e^{-im\phi} + c^\dag_{n,m-1,\sigma} c_{nm\sigma} + h.c.] \\
& - t_{SO} \sum_{nm\sigma \sigma^{'}}[c^\dag_{n,m+1,\sigma}
(i\sigma_x)_{\sigma \sigma^{'}}c_{nm\sigma} \\
& - c^\dag_{n+1,m\sigma} (i\sigma_y)_{\sigma
\sigma^{'}}c_{nm\sigma^{'}} e^{-im\phi} + h.c.] \nonumber
\end{array}
\end{equation}
where $c^\dag_{nm\sigma}$($c_{nm\sigma}$) is the creation
(annihilation) operator for an electron on lattice site $(n,m)$.
$\varepsilon_{nm}$ represents the on-site energy, with magnitude 2t
for 1D chain and 4t for 2D square lattice. Here $t=\hbar^2/2 m^*
a^2$ is the nearest-neighbor hopping energy, which is the energy
scale in this work. Disorder energy $\epsilon_{nm}$, which is
Anderson-type with a uniform distribution in the interval
[-W/2,W/2], is added to the on-site energy $\varepsilon_{nm}$ with W
the disorder strength.

Based on the tight-binding Hamiltonian, the retarded Green's
function in real space is defined as
\begin{eqnarray}
G^r(E)=(E-H-\Sigma^r)^{-1} \nonumber
\end{eqnarray}
where $\Sigma^r$ is the self-energy of the lead, which can be
calculated by a transfer matrix method\cite{transfer}. E is the
electron Fermi energy and chosen to be at the center of the
corresponding subband. By the Fisher-Lee relation\cite{fisher} which
connects the scattering matrix and the Green's function, the delay
time $\tau$ is rewritten in terms of $G^r$ as
\begin{equation}
\tau=-i\hbar Tr[S^{\dag} \frac{\partial S}{\partial E}]=\hbar Tr[G^r
\Gamma G^a] \label{e3}
\end{equation}
where $G^a$ is the advanced Green's function, $G^a=(G^r)^\dag$, and
$\Gamma$ is the line width function describing coupling of the lead
to the scattering region which is given by $\Gamma=i[\Sigma^{r}
-\Sigma^{a}]$. In the presence of large disorders, the system can
either be in diffusive or localized regimes depending on the
dimensionless localization length $\xi/L$. In the localized regime,
the localization length can be defined as\cite{localength}
$<G>=C\exp(-2L/\xi)$ with G the conductance and C is a constant to
be determined. To eliminate C, the localization length $\xi$ at a
particular disorder strength can be obtained from
\begin{equation}
\xi=\frac{2L_2-2L_1}{ln<G_1>-ln<G_2>}.\nonumber
\end{equation}

From Eq.(\ref{e3}), it is clear that the calculation of Wigner delay
time is equivalent to that of density of states. In our numerical
calculation, the real space Green's function can be obtained by
matrix inversion which is very time-consuming. For delay time, one
only needs the first $N_{lead}$ columns of Green's function where
$N_{lead}$ is the dimension of the line width function $\Gamma$. The
transfer matrix method is suitable for this purpose and is fast. To
study the statistics of $\tau$, we need to generate an ensemble of
different realizations of the disordered systems. As we will show in
the next section, the distribution of $\tau$ has a new power-law for
large delay time $\tau>\tau_0$. These large delay times correspond
to rare events with extremely small probability, which means that to
study the new power-law regime of $P(\tau)$ a large configuration
ensemble is necessary. Therefore intensive computation is required
to accumulate enough data for statistical analysis. To speed up the
calculation, we can rewrite the line width function $\Gamma$
as\cite{jian}
\begin{eqnarray}
\Gamma=i[\Sigma^{r} -\Sigma^{a}]=\sum_i^N |W_i\rangle \langle W_i|
\nonumber
\end{eqnarray}
where N is the number of conducting channels in the lead and $|W_i>$
is the renormalized eigenfunction of $\Gamma$. Then the delay time
can be expressed as
\begin{equation}
\tau=\hbar Tr[G^r \Gamma G^a]=\hbar \sum_i^N (G^r|W_i
\rangle)(G^r|W_i \rangle)^{\dag} \label{e5}
\end{equation}
This representation can speed up the numerical calculation and also
enables to calculate the proper delay time $\tau_i$ from the $i-th$
conducting channel. Our results show that using the LU decomposition
with multi-frontal algorithm to solve the linear equation
$(E-H-\Sigma^r)\psi_i = |W_i>$ is faster than the transfer matrix
method especially for large system size like $100 \times 100$.  So
far we have discussed the algorithm to study the statistics of
$\tau$. The numerical results and relevant discussions is the
content of the next section.

\section{numerical results and discussion}

In this section, we will present our numerical results. To test our
code, we first show results of 1D systems with orthogonal symmetry,
in which case the distribution of $\tau$ has been studied
thoroughly. Then in the following two subsections, we shall discuss
in detail the cases of 2D and QD systems with different symmetries
and number of conducting channels.

\subsection{1D tight-binding chain}

For the 1D system, Texier and Comet\cite{texier} derived an analytic
expression of the distribution of Wigner delay time in the localized
regime for high energies or weak disorder strengths. It was found
that $P(\tau)$ has an algebraic tail in the localized
regime\cite{kottos2}
\begin{equation}
P(\tau)=\frac{\xi}{\upsilon \tau^2} e^{-\xi/\upsilon \tau}
\label{e6}
\end{equation}
where $\xi$ is the localization length and $\upsilon$ is the group
velocity. They also found numerically that in the ballistic regime
$P(\tau)$ obeys a Gaussian distribution. The disorder used in their
numerical simulation is the $\delta$ potential with random position
in the system.

\begin{figure}[tbhp]
\centering
\includegraphics[width=8.5cm]{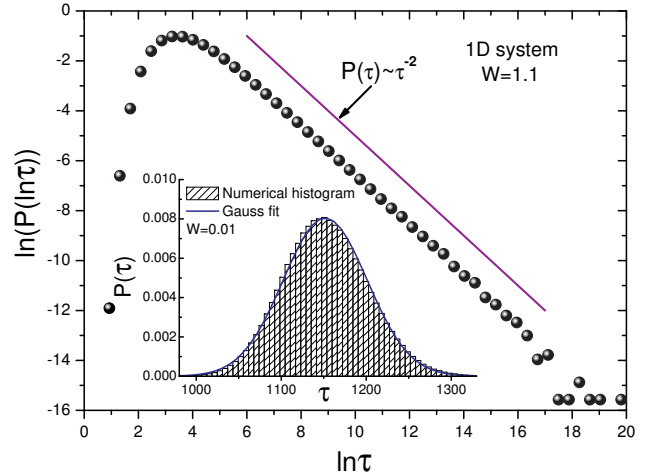}
\caption{The distribution of Wigner delay time of a 1D semi-infinite
long chain at disorder strength $W=1.1$. Inset is the normalized
distribution histogram of $\tau$ at $W=0.01$. Blue curve shows the
Gaussian fitting of the histogram. 4,000,000 configurations have
been generated at each disorder strength W. }\label{fig2}
\end{figure}

We have calculated the delay time distribution for a 1D
tight-binding chain in both the ballistic regime and the localized
regime at a relative high energy $E=1.0$ with disorder strength
$W=0.01$ and $W=1.1$, respectively. The result is shown in
Fig.\ref{fig2}. Clearly the delay time distribution has a Gaussian
shape at weak disorder $W=0.01$ when $L \ll \xi$ (see inset of
Fig.\ref{fig2}). As W increases, $P(\tau)$ transforms gradually from
a symmetric Gaussian distribution to a one-sided
distribution\cite{one-sided}. To make the behavior more transparent,
one may change the variable from $\tau$ to its natural logarithm.
Take Eq.(\ref{e6}) as an example. When we use $\ln \tau$ as the
variable, $P(\ln\tau)=\tau P(\tau)$, and taking logarithm at both
sides, one arrives at
\begin{equation}
\ln P(\ln \tau)=\ln \frac{\xi}{\upsilon} - \ln\tau -
\frac{\xi}{\upsilon} \frac{1}{\tau} \label{e7}
\end{equation}
The first term in the above expression is a constant for a specific
disorder strength and the last term tends to zero when $\tau$ is
very large. As a result, a linear tail arises in the $\ln
P(\ln(\tau))-\ln\tau$ curve. In Fig.\ref{fig2} we see that the
linear tail of $\ln P(\ln(\tau))$ at disorder strength $W=1.1$,
where the system is already localized ($L \gg \xi$), can be well
fitted by a straight line with a slope $-1$, which implies that
$P(\tau) \propto 1/\tau^2$ in the large $\tau$ region. We notice
that there is a departure from the straight line at the end of the
distribution curve. This departure can be improved by using more
configurations, since the larger the $\tau$, the smaller its
occurrence probability. We will explain the origin of these large
$\tau$ in the next subsection.

To summarize briefly, our numerical results for Anderson-disordered
1D system are consistent with the conclusion of Texier and
Comet\cite{texier}, which confirms the universality of the
$1/\tau^2$ power-law tail.

\subsection{2D square lattice}

\begin{figure*}[tbhp]
\centering
\includegraphics[width=18cm]{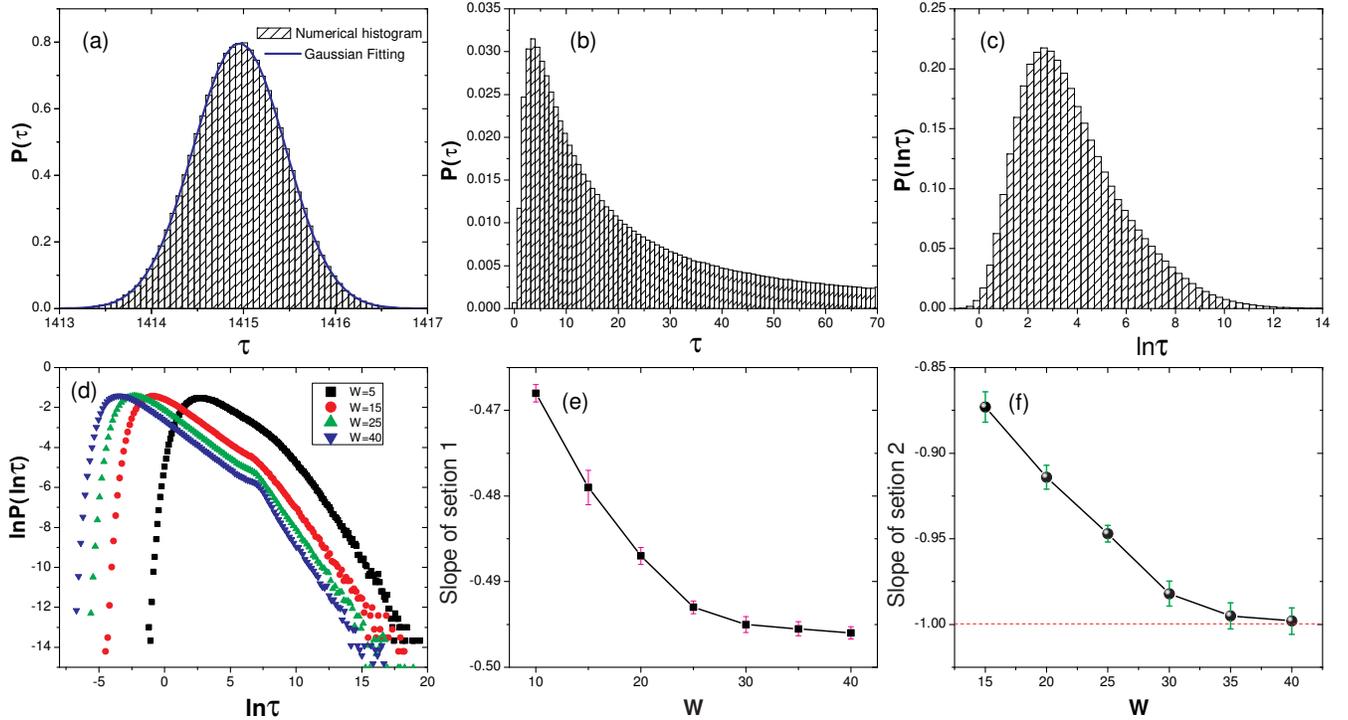}
\caption{Panel (a) is the distribution histogram of $\tau$ at
disorder strength $W=0.0001$ and the blue curve is a standard
Gaussian fitting. Panel (b) is $P(\tau)$ at $W=15$ and the
corresponding histogram with variable changed to $\ln\tau$ is
displayed in panel (c). Panel (d) contains the $\ln P(\ln\tau)$
versus $\ln\tau$ curves for $W=5, 15, 25$, and $40$. Panel (e) and
(f) are respectively the asymptotic behavior of slopes for different
linear sections in panel (d)'s $\ln P(\ln\tau)$ curves. The system
is under such conditions: N=1 and $\beta=1$.}\label{fig3}
\end{figure*}

For 2D systems, most theoretical works of delay time distribution
were within the random matrix theory. As we know, the random matrix
theory works well in the diffusive regime. However, it becomes
difficult to describe properties in the localized regime. In the
diffusive regime, RMT predicts that $P(\tau)$
shows\cite{fyodorov2,sommers} a universal power-law behavior
$\tau^{-1.5}$, which was confirmed by a KR
model\cite{ossipov2,kottos1}. Our numerical results for proper delay
time obtained from 2D Anderson systems also show the $\tau^{-1.5}$
power-law behavior in the intermediate range of $\tau$ and reveal
new properties of the distribution of Wigner delay time in the
localized regime.

First we start with the simplest case of single conducting channel
$N=1$ with preserved time-reversal symmetry ($\beta=1$) and the
results are shown in Fig.\ref{fig3}. Fermi energy of electron is
chosen to be at the center of the first subband, $E=0.004$. From
panel (a) of Fig.\ref{fig3} it is obvious that in the ballistic
regime the distribution of Wigner delay time $\tau$ has a natural
Gaussian shape at a weak disorder $W=0.0001$, which is the same as
that of 1D systems. As the disorder increases, $P(\tau)$ is no
longer symmetrically distributed but spreads over a wide range with
a one-sided peak located at small delay time region, as shown in
panel (b) with $W=15$. From the calculation of localization length
$\xi$, we know that the system is localized at such a disorder
strength. Therefore to get an overview of the distribution including
the long delay time tail, we change the variable from $\tau$ to
$\ln\tau$ and the histogram of $P(\ln\tau)$ is depicted in panel
(c). Based on Eq.(\ref{e7}) and the argument therein, we plot the
logarithm of the distribution of $\ln\tau$ in panel (d) for
different disorder strengths ranging from $W=5$ to $W=40$. We see
that as the disorder is increased in the scattering region, the
curve of $\ln P(\ln\tau)$ vs $\ln\tau$ gradually evolves into a
piece-wise power-law pattern with two different power-laws and an
abrupt change from one to the other at a particular value of
$\tau_0$. This picture becomes clear when W increases to 40. We
understand this behavior as follows. In the scattering system, the
characteristic time scale is set by $\tau_0 = L/v_g$ with $L \sim
3L_0$ where $L_0$ is the dimension of the scattering region and
$v_g$ is the group velocity. In the strong localized regime, there
are two kinds of scattering event naturally separated by $\tau_0$,
one corresponds to the usual direct reflection with delay time
$\tau<\tau_0$ and the other corresponds to multi-resonant reflection
with long delay time $\tau>\tau_0$. For Fig.\ref{fig3} with
$L_0=80$, $\tau_0=3L_0/v_g$ gives $\ln \tau_0 \sim 7.5$ which is
close to the transition point from the power-law $\tau^{-1.5}$ to
the power-law $\tau^{-2}$ in Fig.3. For strong disordered systems,
the large delay time scattering events with $\tau > \tau_0$ are
really rare events with extremely small probability. In addition,
this probability decreases as W increases. For instance, the
probability $P(\tau)|_{\tau > \tau_0}$ is $13\%$ for $W=15$ while
for $W=40$ it drops to $0.32\%$. To study such rare events, more and
more configurations of disordered samples are required for statistic
analysis to get an accurate result. For $W=40$ we have used an
ensemble of over 40,000,000 different realizations.

Another point worth noticing is that the exponents of two power-laws
decrease slowly as disorder strength is increased. These two
exponents versus disorder strength W are plotted in panels (e) and
(f), which converge towards -0.5 and -1.0, respectively. We note
that the curve in panel (f) has larger error bars compared with that
in panel (e). This is understandable since large $\tau$ is more
difficult to sample.

\begin{figure}[tp]
\centering
\includegraphics[width=9cm]{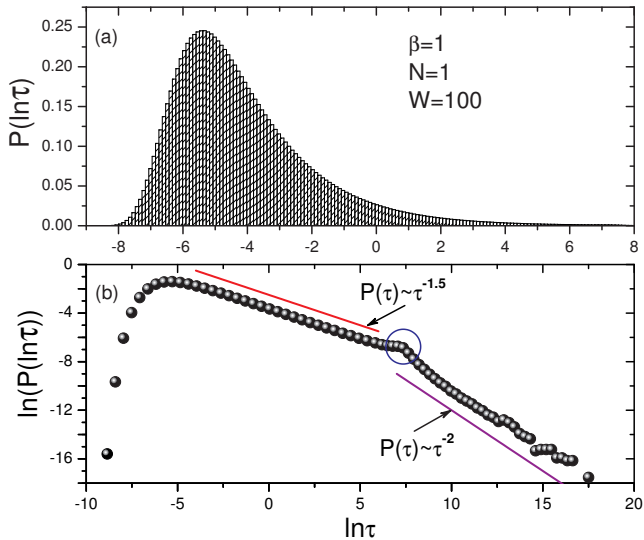}
\caption{Statistics of Wigner delay time at disorder strength
$W=100$ for $N=1$. Panel (a) is the $P(\ln\tau)$ histogram and panel
(b) is the corresponding $\ln P(\ln\tau)-\ln\tau$ curve. Red and
purple straight lines individually have slope -0.5 and
-1.0.}\label{fig4}
\end{figure}

Since the computation is getting extremely time-consuming it is very
difficult if not impossible to obtain the plot shown in panels (e)
and (f) for large disorder strengths. Hence we decide to calculate
$P(\tau)$ for a particular disorder strength which is large enough
to determine the exponents for both power-law behaviors. After some
trial and error, we found the disorder strength $W=100$ is
appropriate for this purpose. The exponent of power-law in the
intermediate region is already converged at such a W and does not
change upon further increasing of disorder strength to $W=200$. To
get a clear view of the tail with small probability, a large
ensemble of 420,000,000 configurations is accumulated and the
analyzed data is shown in Fig.\ref{fig4}. Clearly the algebraic tail
where $\tau > \tau_0$ is invisible in the histogram of panel (a),
since it accounts for only 0.118 \% in the ensemble. From panel (b)
one clearly sees that the first power-law of $\ln P(\ln\tau)$ has a
slope of -0.5, corresponding to a power law distribution $P(\tau)
\sim \tau^{-1.5}$. After a transition region around $\tau=\tau_0$
shown by the blue circle in Fig.\ref{fig4}, the second power-law
distribution is found to have a slope of -1, which means that it is
a power-law tail $P(\tau) \sim \tau^{-2}$. The fluctuation at the
end of the tail arises because there is not enough configurations.
In such a strong localized regime with $W=100$, delay time with
magnitude $\ln \tau > 15$ often has a few configurations out of the
total 420,000,000 ensemble.

The power-law distribution $\tau^{-1.5}$ has already been predicted
by RMT\cite{fyodorov2,sommers} and was also confirmed numerically
using KR model\cite{kottos2} in the diffusive regime. Our results
show that this power law $\tau^{-1.5}$ for {\it each proper delay
time} exists for any $N$ where $N$ is the number of conducting
channels (also see numerical results presented below). In addition,
this behavior persists in the strong localized regime. Importantly,
the $\tau^{-2}$ algebraic tail is a novel feature that has not been
reported before in 2D systems. The physical origin of the novel
power-law behavior is due to the Azbel resonance states or necklace
states\cite{azbel1,azbel2,necklace1}, which are extended states in
localized systems. Since in the localized regime most of the
electron states are localized, these necklace states or Azbel
resonances survive through multiple-resonances. These states are
really rare events in the disordered samples so that they have an
extremely small probability to occur. For example, in the above
calculation with N=1 and $W=100$, the probability for those states
with $\tau > \exp(8)$ is $0.03 \%$. Despite of their rare nature,
these states can play a significant role in the distribution
function of delay time and dominate in the density of states in
disordered systems. It is because a necklace state experiences
multiple resonant scattering giving rise to a large density of
states. For instance, emittance calculations in disordered 1D, 2D,
and QD systems show that due to the necklace states, the average
emittance remains negative in the localized regime.\cite{emittance}
In fact, the existence of necklace states has already been observed
through optical experiment\cite{necklace2} in a quasi-1D system. The
fact that the signature of necklace states is observed again in
disordered 2D systems with single lead indicates that the
non-localized necklace states are generic in strong disordered
systems, although difficult to see them.

\begin{figure}[thp]
\centering
\includegraphics[width=9cm]{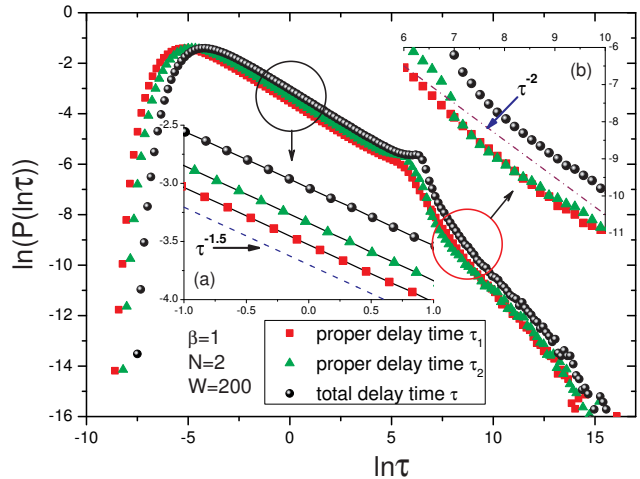}
\caption{Distributions of proper delay time $\tau_1$ and $\tau_2$
and the total $\tau$ at disorder strength $W=100$ in the case of 2
conducting channels. Inset panel (a) highlights the first linear
section of distribution with a -0.5 slope (blue-dash straight line),
while panel (b) for the algebraic tail with a -1 slope
(purple-dash-dot-dash straight line). Ensemble size is 150,000,000.
}\label{fig5}
\end{figure}

Next we consider the case of two conducting channels ($N=2$) in a
system with $\beta=1$. Here electron Fermi energy is fixed at the
center of the second subband, where $E=0.01$. Since the group
velocity of electron incident from different subbands are different,
proper delay times for different conducting channels are not
statistically equivalent. Hence the properties of the proper delay
time $\tau_i$, which describes the scattering of electron of the
$i^{th}$ channel, as well as the total Wigner delay time
$\tau={\sum_i}^N \tau_i$ are studied separately. To save
computational time, we calculated the distribution of relevant times
only at a large disorder strength $W=100$. As we have seen above, at
this $W$ $P(\tau)$ of $N=1$ converges to the piece-wise power-law
($\tau^{-1.5}$ and $\tau^{-2}$) behavior. Statistically analyzed
result upon an ensemble of 150,000,000 configurations is shown in
Fig.\ref{fig5}. To separate the curves we zoom in the picture around
two particular areas of the two power-law regions, which are
highlighted in the insets.

\begin{figure}[tbhp]
\centering
\includegraphics[width=9cm]{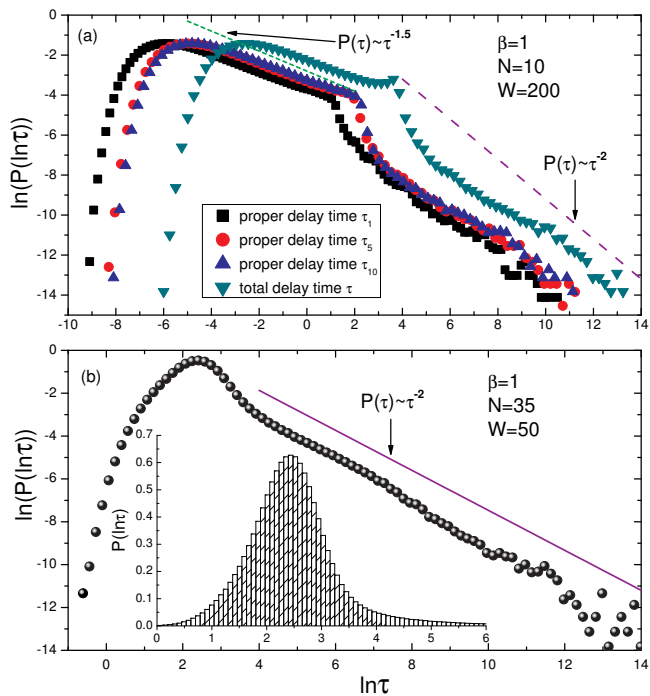}
\caption{Panel (a) shows the distribution of proper delay time
$\tau_1$, $\tau_5$, and $\tau_{10}$ as well as the total delay time
$\tau$ at $W=200$ with 10 conducting channels in the lead. Panel (b)
contains the normalized distribution histogram and its logarithm
correspondence of total delay time $\tau$ at $W=50$ for
$N=35$.}\label{fig6}
\end{figure}

From Fig.\ref{fig5} it is clear that proper delay time $\tau_1$ and
$\tau_2$ follow the same distribution with different constants
($P(\tau_{1/2}) \sim C_{1/2} \tau^{\alpha}$ with $\alpha=$  -1.5 or
-2). It shows that the proper delay time of the first and second
subbands are statistically independent. Since the total delay time
$\tau$ is the summation of $\tau_1$ and $\tau_2$, it is not
surprising that the distribution of total delay time $\tau$ keeps
the essential feature of individual proper delay time. Clearly all
three curves show piece-wise power-law behaviors, which is similar
to the case of a single channel $N=1$ (see Fig.\ref{fig4}). In view
of the distribution for systems with $N=1$ and $N=2$, we expect that
the delay time distribution in multichannel case ($N \geq 2$) also
show piece-wise power-law behavior. However the transition region in
the distribution curve of $\tau$ is broadened due to the overlap of
$P(\tau_1)$ and $P(\tau_2)$. As the channel number increases, this
may lead to a significant change of the distribution form as will be
discussed in detail below.

Numerical results for cases of more conducting channels $N=10$ and
35 within orthogonal ensemble ($\beta=1$) are shown in
Fig.\ref{fig6}. Following observations are in order. (1). For a
strong disorder strength W=200, statistical analysis on the
distribution of proper delay time at $N=10$ shows that the
individual $\tau_i$ exhibits a piece-wise power-law behavior with
two scalings $\tau^{-1.5}$ and $\tau^{-2}$, similar to that of the
proper delay time in $N=2$ case. In addition, the total delay time
distribution $P(\tau)$ at $N=10$ given by the superposition of all
the $P(\tau_i)$ (i=1,2,.., 10) results in the same behavior (see
panel (a)). We also note that due to the group velocity difference
for different subbands, the transition point $\tau_0 = 3L_0/v_g$ of
proper delay time distribution is not at the same position in $\ln
P(\ln\tau)$-$\ln\tau$ curve. The transition value of proper delay
time is the smallest for the first subband while the largest for the
highest subband. The overlap of proper delay time distribution
causes the broadening of transition region of the distribution
$P(\tau)$. (2). Comparing the situations of $N=2$ and $N=10$
(Fig.\ref{fig5} and panel (a) of Fig.\ref{fig6}), the broadening is
more obvious at large N. When the number of conducting channel
increases to $N=35$, the broadening is so significant that it
destroys the power-law behavior of $\tau^{-1.5}$ in the total delay
time distribution although $\tau^{-1.5}$ power-law exists for each
proper delay time distribution. Therefore our results suggest that
in the large N limit, only the power-law of $\tau^{-2}$ survives.

\begin{figure*}[tp]
\centering
\includegraphics[width=16cm]{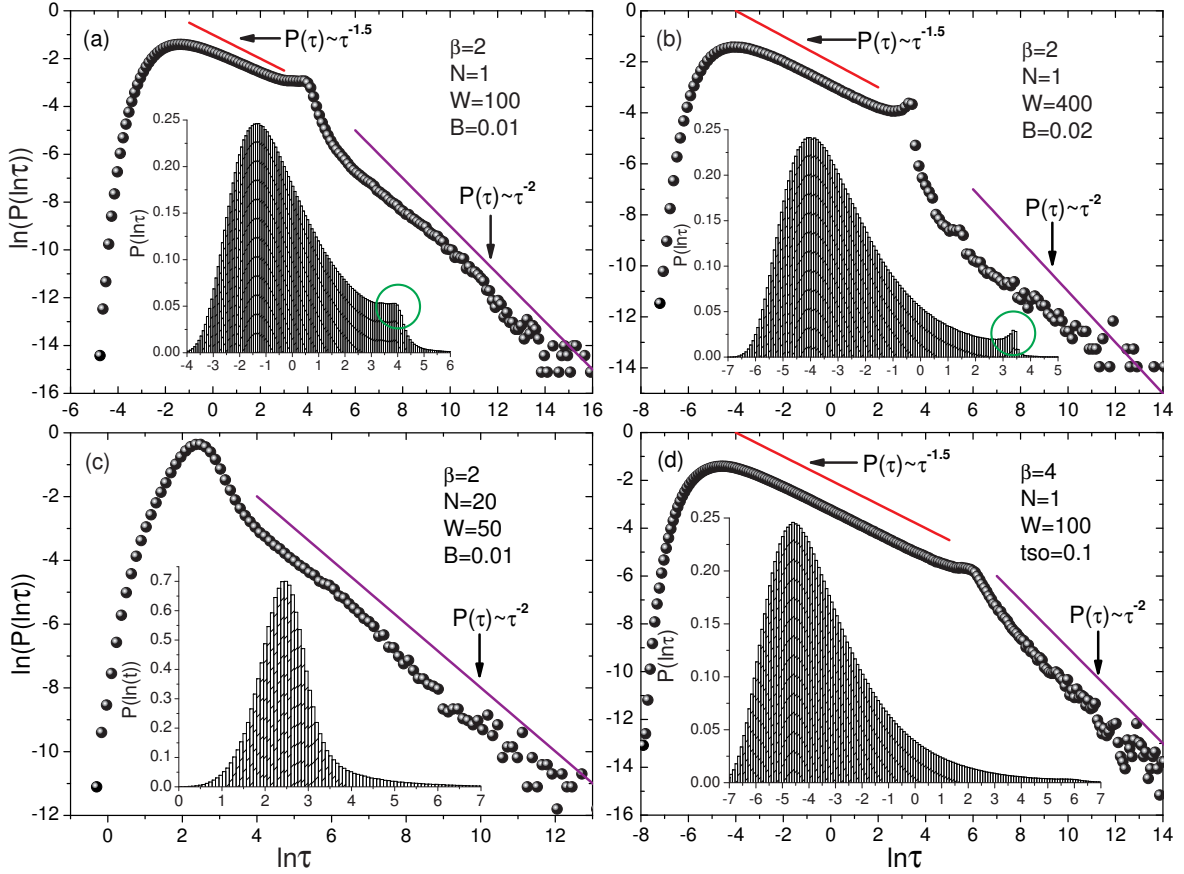}
\caption{Panel (a) and (b) are respectively the distributions of
delay time at two different sets of parameter. In panel (a) it is
$W=100$ and $B=0.01$, while in (b) $W=400$ and $B=0.02$. Both
systems have single conducting channel($N=1$) in unitary symmetry
case ($\beta=2$). Ensemble size is 63,000,000. Panel (c) shows the
statistics of $\tau$ at multi-channel case with $N=20$ at $\beta=2$.
Case of symplectic symmetry case ($\beta=4$) is drawn in panel (d)
with $W=100$and $N=1$, where an ensemble of size 120,000,000
configurations is used.}\label{fig7}
\end{figure*}

\begin{figure*}[tp]
\centering
\includegraphics[width=16cm]{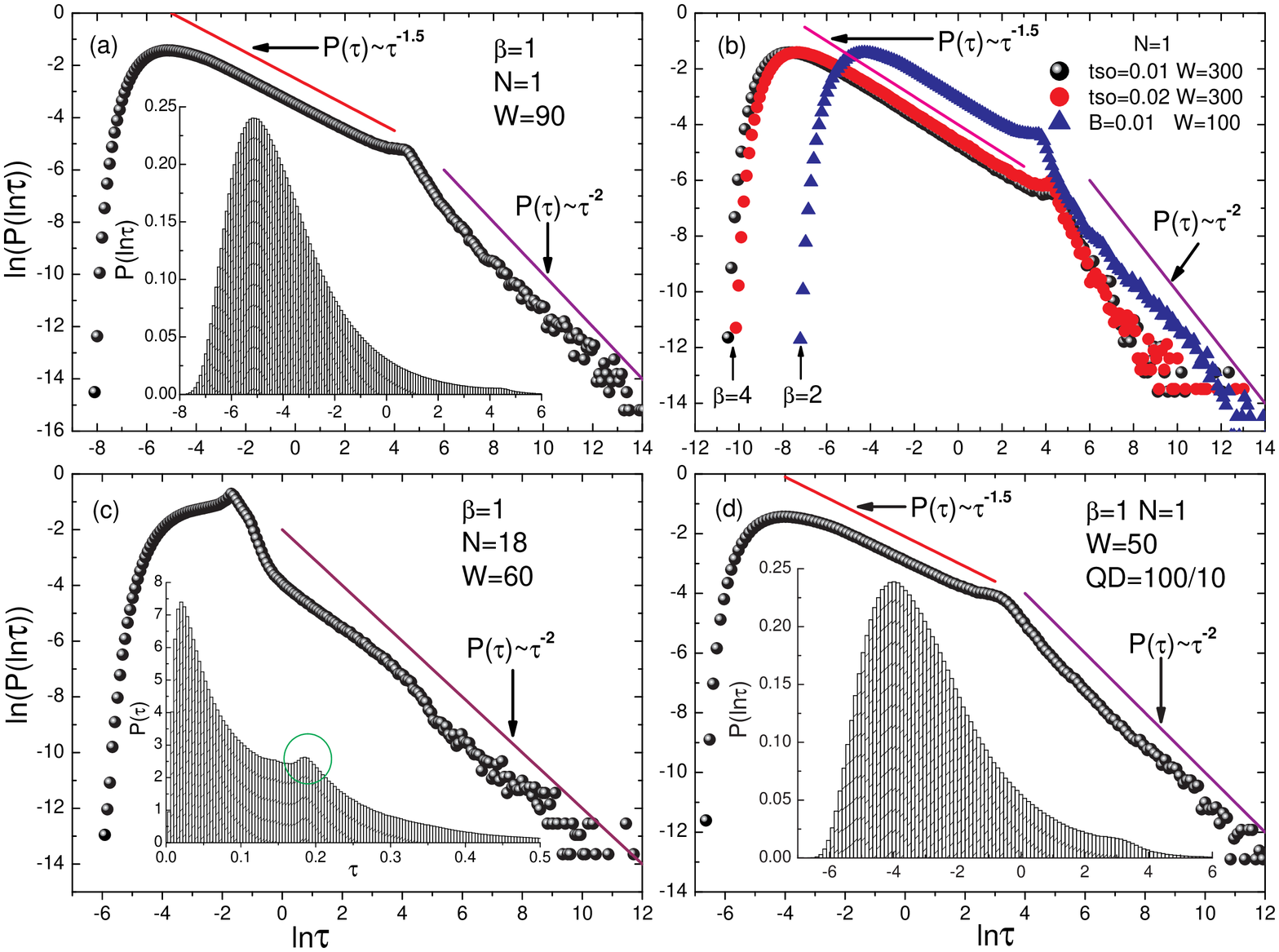}
\caption{Quantum dot case. Panel (a) and (b) are the distribution of
delay time for different symmetry classes ($\beta=1, 2$, and 4) with
single conducting channel($N=1$). Panel (c) corresponds to $N=18$ in
orthogonal symmetry $\beta=1$. Calculation parameters are shown in
the graphs. The result of a large QD at $\beta$=1 is presented in
panel (d). }\label{fig8}
\end{figure*}

Now we turn to the systems with symmetry class $\beta=2$ where the
time reversal symmetry is broken and $\beta=4$ in the absence of
spin-rotation symmetry. In our numerical calculation, the unitary
symmetry class ($\beta=2$) is realized by applying an external
magnetic field while for symplectic symmetry ($\beta=4$) we consider
the spin-orbit interaction. In both unitary and symplectic cases we
mainly numerically investigate the statistics of Wigner delay time
for single conducting channel ($N=1$) since the computation is
getting extremely time-consuming for 2D with $\beta=2$ and 4. From
the discussion in orthogonal symmetry case, some intuition can be
obtained for the distributions of delay time at multi-channel case.
From Fig.\ref{fig7} we see that there are clearly two power-law
regions with exponents approaching $-1.5$ and $-2$ for total delay
time distribution when N=1 (panels (a) and (b)). We have tested two
different magnetic field strengths and they give similar results. In
addition, the transition phenomena from one power-law to another can
also be seen from inset of Fig.\ref{fig7}(a) and (b) that are
different from the case with $\beta=1$. We have checked that the
proper delay time distribution follows a piece-wise power-law
behavior. From our experience in orthogonal ensemble we expect that
the $\tau^{-1.5}$ intermediate region will be destroyed at large N.
Indeed, our results confirm this expectation (see panel (c) where
$N=20$ is investigated). Except for this difference shown by green
circles in Fig.\ref{fig7}, the delay time distribution is similar to
that of the system with reserved time-reversal symmetry
($\beta=1$)(panel (a) of Fig.\ref{fig4}). Finally, in the case of
symplectic symmetry, the distribution is the similar to that of
$\beta$=1. We have also varied the strength of spin-orbit coupling
$t_{SO}$ and found that it does not affect the distribution
behavior.

To briefly summarize our results for 2D systems, the Wigner delay
time distribution in the 2D lattice system shows a piece-wise
power-law feature in strong localized regime, with power-law
converging to $\tau^{-1.5}$ in the intermediate region and the
algebraic tail obeying a different power law $\tau^{-2}$ for a
single conducting channel. The two power-law regions are separated
by a clear transition region determined by $\tau_0$ in the
distribution curve. For multi-conducting channels, the distribution
of each proper delay time is found to have the same power-law
distribution as that of the delay time $\tau$ for $N=1$. For a few
conducting channels, the distribution of total Wigner delay time
behaves like that of a single channel. When N is large, however, the
superposition of each proper delay time distribution gives rise to a
broad transition region that gradually destroys the $\tau^{-1.5}$
power-law region. In the large N limit, only one power-law region
exists which corresponds to $\tau^{-2}$. These features seem to be
independent of symmetry class of 2D systems, i.e., regardless of
orthogonal, unitary or symplectic symmetries.

\subsection{Quantum dot system}

The quantum dot (QD) system consists of a square scattering region
and single lead attached at the right side with 1/4 width of the
scattering region, as shown in Fig.\ref{fig1}. Numerical results
depicted in Fig.8 show that the conclusions obtained in 2D system
are also applicable to QD case. In Fig.\ref{fig8}(a) and (b), we see
that the piece-wise power-law behavior is observed again in all
Dyson's symmetry classes for the case of $N=1$. For $\beta=1$ and
$N=18$, we see that the power-law $\tau^{-1.5}$ no longer exists
that is similar to the situation in 2D systems. A slight difference
compared to the 2D case is that there is an additional peak in the
distribution of delay time $P(\tau)$ vs $\tau$ that indicates the
onset of the power-law region of $\tau^{-2}$(see panel (c) of
Fig.\ref{fig8}). To check the size effect of the distribution, we
have examined a larger quantum dot system with size of $L=100$ and
lead width $W_0=10$. The numerical results for $\beta=1$ and $N=1$
are shown in Fig.8(d) and similar conclusion can be drawn.

\section{conclusion}

In conclusion, we numerically investigate the statistical properties
of Wigner delay time in Anderson disordered 1D, 2D and quantum dot
systems for different symmetry classes $\beta=1,2,4$. The proper
delay time distribution is found to be universal in strong localized
regime for 2D and QD systems and shows a piece-wise power-law
behavior. In addition to the known power-law scaling region
${\tau^{-1.5}}$, a novel power-law region of ${\tau^{-2}}$ is
identified in the localized regime, which is independent of the
number of conducting channel N and Dyson's symmetry classes $\beta$.
Our results indicate that the existence of necklace states is
responsible for the $\tau^{-2}$ algebraic tail, which are rare
events and have an extremely small distribution probability. The
total delay time distribution can behave differently. For a few
conducting channel, there is a crossover region from one power-law
region to another. As the number of conducting channels N increases,
this crossover region broadens and the power-law region of
${\tau^{-1.5}}$ becomes narrow. In the large N limit, the power-law
of ${\tau^{-1.5}}$ is destroyed due to the broadening of the
crossover region and only the $\tau^{-2}$ algebraic tail survives.

\section{acknowledgments}

This work is supported by RGC grant (HKU 705409P) from the HKSAR and
LuXin Energy Group. We thank the HPC POWER of the computer center,
HKU for computation resources allocation.

\end{document}